# The Computational Complexity of Truthfulness in Combinatorial Auctions

Shahar Dobzinski    Jan Vondrák

September 3, 2018

**Abstract**

One of the fundamental questions of Algorithmic Mechanism Design is whether there exists an inherent clash between truthfulness and computational tractability: in particular, whether polynomial-time truthful mechanisms for combinatorial auctions are provably weaker in terms of approximation ratio than non-truthful ones. This question was very recently answered for universally truthful mechanisms for combinatorial auctions [4], and even for truthful-in-expectation mechanisms [12]. However, both of these results are based on information-theoretic arguments for valuations given by a value oracle, and leave open the possibility of polynomial-time truthful mechanisms for succinctly described classes of valuations.

This paper is the first to prove *computational hardness* results for truthful mechanisms for combinatorial auctions with succinctly described valuations. We prove that there is a class of succinctly represented submodular valuations for which no deterministic truthful mechanism provides an $m^{1/2-\epsilon}$-approximation for a constant $\epsilon > 0$, unless $NP = RP$ ($m$ denotes the number of items). Furthermore, we prove that even truthful-in-expectation mechanisms cannot approximate combinatorial auctions with certain succinctly described submodular valuations better than within $n^\gamma$, where $n$ is the number of bidders and $\gamma > 0$ some absolute constant, unless $NP \subseteq P/poly$. In addition, we prove computational hardness results for two related problems.

## 1 Introduction

Is it possible to design truthful polynomial-time mechanisms that achieve a good approximation ratio? This is one of the basic questions in Algorithmic Mechanism Design. The focus of most of the work on this question is on settings in which on one hand there exists a truthful algorithm that maximizes the social welfare (VCG), but on the other hand computing the optimal solution is computationally intractable. The goal is therefore to determine whether there exist truthful polynomial time mechanisms that guarantee good approximation ratios. Several settings were considered in the literature, but the flagship challenge is to design polynomial-time truthful approximation mechanisms for the problem of combinatorial auctions.

In a combinatorial auction, we want to sell $m$ items to $n$ bidders with valuation functions $v_i : 2^{[m]} \to \mathbb{R}_+$. As usual, we assume that $v_i(\emptyset) = 0$ and that $v_i$ is monotone, i.e. $v_i(S) \leq v_i(T)$ whenever $S \subset T$. The goal is to design a mechanism that allocates disjoint sets $S_1, \ldots, S_n$ to the $n$ bidders, optimizing (at least approximately) the *social welfare* $\sum_{i=1}^n v_i(S_i)$, in a way that incentivizes the bidders to report their true valuations (the property of *incentive-compatibility*, or *truthfulness*). This is done by charging *payments* $p_1, \ldots, p_n$ by the mechanism, so that for each player, reporting his true valuation maximizes the profit $v_i(S_i) - p_i$ (in the case of truthfulness in expectation, the reporting the true valuation maximizes the expectation of this expression).

Without the requirement of truthfulness, combinatorial auctions admit constant-factor approximation algorithms for various non-trivial classes of valuations functions, in particular for *submodular valuations*: A valuation $v$ is called submodular if for each $S$ and $T$ we have that $v(S) + v(T) \geq v(S \cap T) + v(S \cup T)$. Combinatorial auctions with submodular valuations admit a $(1-1/e)$-approximation [21] and it is also known that this approximation is optimal [16]. On the other hand, the VCG mechanism is truthful and provides optimal social welfare, but naturally is not computationally efficient. Therefore, submodular valuations form a natural setting to investigate the question whether truthfulness can be reconciled with polynomial-time approximation or not.

The best known polynomial time truthful mechanism for combinatorial auctions with submodular valuations achieves a factor of $O(\sqrt{m})$ [8]. This is quite a poor ratio indeed, but there was hope that one can significantly improve over it. Indeed, for a variety of other interesting settings good truthful mechanisms do exist. For example, if bidders still have submodular valuations but are able to answer the more complicated demand queries, a randomized truthful $O(\log m \log \log m)$-approximation mechanism exists [3], a powerful construction of [18] yields optimal truthful-in-expectation mechanisms for many settings, and very recently [11] gave VCG-based truthful-in-expectation $(1-1/e)$-approximation mechanisms for combinatorial auctions with explicit coverage valuations (and more general valuations in a certain oracle model). This and related results raised hopes that truthfulness (or at least truthfulness-in-expectation) might be reconcilable with computational efficiency.

On the negative side, it was proved first that any universally truthful VCG-based (a.k.a. maximal in range) mechanism that achieves an $m^{1/6-\epsilon}$-approximation for submodular valuations requires exponential communication [6]. Still, this result did not rule out the possibility of a different, non-VCG mechanism, and it also did not address the possibility of truthful-in-expectation mechanisms. Only recently, [4] proved that if a deterministic truthful or universally truthful mechanism can access the valuations only through *value queries* (given $S$, what is $v(S)$?) and guarantees an approximation ratio of $m^{1/2-\epsilon}$, for any constant $\epsilon > 0$, then it must make exponentially many value queries. Moreover, [12] proved that even truthful-in-expectation mechanisms that provide an approximation ratio better than $m^\gamma$, for some constant $\gamma > 0$, require exponentially many value queries.

The skeptical reader may wonder now whether these results [4, 12] are a symptom of the value oracle model, and whether they just reflect an informational bottleneck: value oracles are too weak for transferring sufficient information from the bidders to the mechanism. One may further argue that when bidders are required to provide a succinct explicit description of their valuation, in some agreed-upon bidding language, these informational bottlenecks disappear. Consequently, good computationally efficient mechanisms may exist.

In some domains, like multi-unit auctions where all items are allocated [17] and combinatorial public projects [20], the hardness of *deterministic* truthful polynomial time algorithms for succinct valuations with good approximation ratios is already known. However, their technique does not seem to be applicable to combinatorial auctions or any other auction domain. The reason is that both papers use the characterize-and-optimize approach: they first characterize the truthful mechanisms in the domain, showing them to be VCG-based, and then show that computationally-efficient VCG-based mechanisms do not provide good approximation ratios, under plausible complexity assumptions. The characterization step in these papers heavily relies on the fact that the domains exhibit strong forms of externalities between the bidders. Combinatorial auctions, and auction domains in general, are different since the value of a bidder depends only on the bundle he receives. Recently a characterization of "scalable" two-player multi-unit auctions was obtained in [7], but even there the additional assumption of scalability is needed, and it is not clear if we can extend it to more than two players and submodular valuations. Furthermore, no characterizations of truthful-in-expectation mechanisms are known at all, and such characterizations are currently beyond the reach of our techniques, even for relatively simple domains that do exhibit strong externalities.

The only negative result for combinatorial auctions with succinct valuations so far has been [1]. There, building on the work of [20], it is proved that any universally truthful VCG-based mechanism that achieves a $m^{1/2-\epsilon}$-approximation for budget-additive valuations (and some other natural, succinctly represented classes of valuations) would imply $NP \subseteq P/poly$ [1]. But just as previously in the information-theoretic setting, the question of the existence of a non-VCG based mechanism, or even a VCG-based truthful-in-expectation (MIDR) mechanism, remained open. Recall that such mechanisms are known to give non-trivial results - see above. Thus the question still remained, whether truthful mechanisms for combinatorial auctions with succinctly represented valuations are in principle less powerful than non-truthful ones.

---
[1]For succinct representations, the only barrier to designing a truthful mechanism is computational complexity, since with unbounded computational power, the truthful VCG mechanism provides an optimal solution. Thus any negative results in this setting will likely depend on complexity-theoretic assumptions.

In this paper we give a definite answer to this question, and show that unless $NP \subseteq P/poly$, there are no truthful polynomial-time constant-factor approximation mechanisms for combinatorial auctions with succinctly represented submodular valuations (while there is a known $(1 - 1/e)$-approximation for such valuations without truthfulness). This result rules out even truthful-in-expectation mechanisms. This is the first separation in terms of *computational complexity* between truthful and non-truthful mechanisms for combinatorial auctions. For deterministic mechanisms, our result is even stronger: we prove a tight $m^{1/2-\epsilon}$ hardness factor. In other words, even if able to bypass the information bottleneck, the auctioneer will face an equally hard barrier to pass: the computational one.

**Our main results** All of our results hold for succinctly represented valuations, in particular succinctly represented *submodular* valuations. We say that a class $\mathcal{C}$ of valuation functions $v : 2^{[m]} \to \mathbb{R}_+$ is succinctly represented by an encoding $\nu$, if for each $v \in \mathcal{C}$, $\nu(v)$ is a bit string of size polynomial in $m$ and there is a procedure that, given $\nu(v)$ and $S \subseteq [m]$, computes $v(S)$ in time polynomial in $m$. Examples of succinctly represented functions that where studied in the literature include budget additive valuations, coverage valuations, and XOS (with polynomially many clauses). See, e.g., [19].

Our two main results are for combinatorial auctions:

- We prove that there is a class of succinctly represented submodular valuations for which no deterministic truthful mechanism provides an $m^{1/2-\epsilon}$-approximation for a constant $\epsilon > 0$, unless $NP = RP$.

- We prove that there is a constant $\gamma > 0$ such that for any constant number of bidders $n$, there is a class of succinctly represented submodular valuations for which no randomized truthful-in-expectation mechanism achieves a $n^\gamma$-approximation, unless $NP \subseteq P/poly$.

**Results for other domains** We complement our main results with two additional impossibilities for other domains: combinatorial public projects and multi-unit auctions.

In a *combinatorial public project*, a single set $S$ should be chosen for $n$ bidders with valuation functions $v_i$, maximizing $\sum_{i=1}^n v_i(S)$ under the constraint that $|S| = k$.

- We prove that for combinatorial public projects with coverage valuations, there is no constant-factor truthful-in-expectation mechanism, unless $NP \subseteq P/poly$.

We remark that it was already known that combinatorial public projects with coverage valuations do not admit any *deterministic* truthful mechanism with approximation better than $m^{1/2-\epsilon}$ [2]. On the other hand, the problem of *flexible* combinatorial public projects, where the constraint $|S| = k$ is replaced by $|S| \leq k$, does admit a truthful-in-expectation $(1 - 1/e)$-approximation for coverage valuations [10]. Thus we prove a separation between flexible ($|S| \leq k$) and exact ($|S| = k$) combinatorial public projects.

We prove a related separation in the domain of *multi-unit auctions*. In [5] it is shown that there exists a polynomial time truthful-in-expectation FPTAS for multi-unit auctions. This mechanism is *maximal in distributional range* (MIDR). It is also shown there that no polynomial time truthful-in-expectation mechanism for two players in the oracle model can provide an approximation ratio better than 2, if all items are allocated. We complement this result by showing that no polynomial-time MIDR mechanism for two players with succintly described valuations achieves an approximation ratio better than 2, unless $RP = NP$.

**A word on our techniques** This paper builds on advances developed in a sequence of recent papers [4, 12, 9]. Our proofs for combinatorial auctions are obtained by considering the previously developed proofs in the value oracle model [4, 12], and converting them into computational hardness proofs in the following manner: The proofs in [4, 12] depend on a certain "hidden set" which is hard to find by value queries. In our new proofs, the hidden set in both cases is represented implicitly as a solution to some computationally difficult problem (e.g. SAT). However, some new difficulties arise when trying incorporate this idea in the proof of hardness for truthful mechanisms.

In the case of deterministic truthful mechanisms, the construction of the valuation function in [4] itself depends on the properties of the presumed mechanism, namely on its pricing scheme. We resolve this issue by incorporating a component in the representation of a valuation that allows us to feed a description of the presumed mechanism itself into the valuation, thus obtaining a contradiction by a diagonalization argument.

In the case of truthful-in-expectation mechanisms, the location of the hidden set affects the valuation function more substantially than just the value of a single set. We need a more sophisticated encoding to be able to evaluate the function efficiently. We appeal to the encoding of submodular functions using *list-decodable codes* which was used recently in [9] to obtain inapproximability results for submodular optimization problems. Using list-decodable codes, we are able to encode our valuations in a way that simultaneously allows the flexibility of a "hidden set" described implicitly by a computationally difficult problem, and efficient evaluation. An additional technical difficulty in this proof is the presence of certain parameters necessary for a reduction, whose existence is proved non-constructively in [12]. We have to make sure that these parameters can be described by a polynomial-size advice string (leading to the conclusion that $NP \subseteq P/poly$).

Our hardness result for combinatorial public projects is more direct, appealing to the technical properties of Feige's reduction for the hardness of Max-$k$-cover [14]. We use the known fact that combinatorial public projects with 1 bidder essentially have to maximize over a certain range of distributions [12], and using the structure of Feige's reduction, we obtain a contradiction with the hardness of finding good solutions for the Max-$k$-cover problem.

An interesting feature of all our proofs is that they do not rely on the Sauer-Shelah lemma or VC-dimension, which played an important role in previous computational hardness results for combinatorial public projects and combinatorial auctions [20, 1, 2]. One reason for this is that in the case of truthful-in-expectation mechanisms, it is not clear how VC-dimension should be used at all - the range of possible distributions output by the mechanism is a continuum which does not lend itself to combinatorial analysis such as VC-dimension. But even in our proof for deterministic mechanisms, we do not appeal to the Sauer-Shelah lemma. Rather, we design the representation of a valuation function in a way that allows the embedding of a computationally hard problem, and yet admits efficient evaluation of the function.

**Open questions** We leave several interesting questions for future work. First, in the bidding languages we use to prove the hardness of combinatorial auctions, answering demand queries is NP-hard. Can one prove an analogous hardness result for bidding languages in which demand queries can be answered efficiently? This seems to be a challenging question that requires substantial extensions of our techniques. For example, with demand queries it is possible to achieve a logarithmic approximation via a universally truthful mechanism, which beats the bounds we give. Another question is to prove impossibility results for specific bidding languages, like budget-additive, coverage, and XOS.

## 2 Deterministic Truthful Mechanisms for Combinatorial Auctions

Our first proof is about the impossibility of deterministic truthful mechanisms for combinatorial auctions with submodular bidders. In [4] it is shown that any deterministic truthful mechanism for combinatorial auctions with submodular bidders that provides an approximation ratio better than $m^{\frac{1}{2}}$ makes exponentially many value queries. This is proven using a direct hardness approach. The following theorem (non trivially) extends this approach and shows how to obtain impossibilities that are based on computational complexity.

**Theorem 2.1.** *There is a class $\mathcal{C}$ of succinctly represented monotone submodular valuations, such that if there is a deterministic truthful polynomial-time mechanism for combinatorial auctions with valuations in $\mathcal{C}$ that guarantees an approximation ratio of $m^{1/2-\epsilon}$, for any fixed $\epsilon > 0$, then $NP = RP$.*

**Proof sketch** We show that if a truthful algorithm provides a good approximation, then there must exist a player $i$ and valuations $v_{-i}$ such that there exists a submenu of the menu induced by $v_{-i}$ that is not only large but also has a some specific structure. Then we show that for certain valuations of player $i$, finding a profit maximizing bundle (which is a must for a truthful mechanism according to the taxation principle) is computationally hard, and allows one to solve the SAT problem by reduction.

The first part of the reduction is to find the valuations $v_{-i}$ that induce a large, structured menu for player $i$. This is done by showing that a set of random "polar additive" valuations induce a menu with the required structure with high probability. The second part is the optimization part, in which we present valuations $v_i$ for which finding the bundle that maximizes the profit is computationally hard. In [4] this is done by constructing a generic family of valuations using the properties of the structured menu induced by $v_{-i}$, and choosing a specific valuation by giving a bonus to one particular bundle (the intended profit maximizing bundle). However, in this paper we work with succinctly described valuations, i.e. each valuation should have a polynomial-size description which allows one to evaluate the function efficiently. This causes an extra layer of complication: not only that the structured menu depends on the truthful mechanism itself and the valuations of the other players, we cannot even explicitly describe it, since is exponentially large. To overcome this obstacle, we use a "diagonalization argument": the description of a valuation $v_i$ involves the description of an algorithm (or a circuit) that verifies whether a given set $S$ is in the structured menu and computes the correct value $v_i(S)$. Eventually, we feed a circuit describing the presumed mechanism itself into the valuation.

We also have to specify succinctly the special profit-maximizing bundle, so that finding it is computationally hard. We do that in two steps. First, we choose some random projection of every bundle in the structured submenu to $\{0,1\}^\ell$, where $\ell = poly\{n,m\}$. Next, we would like to specify some special strings in $\{0,1\}^\ell$ such that bundles in the structured menu that are projected to the special strings will be the profit-maximizing bundles. Of course, if we just specify the special strings in the straightforward way, it might be easy for an algorithm to efficiently find the profit-maximizing bundle. Therefore, we specify the special strings to be the satisfying assignments of a SAT instance, which completes our reduction.

To summarize, we randomly construct valuations $v_{-i}$, and we show that with some constant probability the $v_{-i}$ induce an exponentially large structured submenu. We then construct a valuation $v_i$ of the $i$'th player based on this structured submenu; we use a decription of the mechanism itself to do so. The profit-maximizing bundles of $v_i$ will be exactly all bundles that are (randomly) projected to a satisfying assignment of the formula. We will show that with inverse-polynomial probability the construction of the random valuations succeeds and the random projection indeed maps a bundle in the menu to a satisfying assignment, and in this case a truthful algorithm must find a solution to the SAT instance. This will show that the existence of a truthful mechanism as in the theorem implies $NP = RP$.

## 2.1 The Bidding Language

We start the proof by specifying the bidding language that we use, i.e. the class of valuations and their representation on the input.

**Additive valuations** Our class of valuations includes the set of all additive valuations: for each bundle $S$, $v(S) = \Sigma_{j \in S} v(\{j\})$. We may use any natural representation of additive valuation to do that. In particular, the proof uses a certain type of additive valuations called *polar additive* valuations. These are additive valuations such that for each item $j$ either $v(\{j\}) = 1$ or $v(\{j\}) = \frac{1}{m^3}$. We sometimes use *random polar additive valuations*. As the name suggests, these are polar additive valuations in which the value of each item $j$ is set independently at random $v_i(j) = 1$ with probability $p = \frac{1}{n}$, or $v_i(j) = \frac{1}{m^3}$ with probability $1 - p$.

**Bonus valuations** In addition, the bidding language is able to express the following *bonus valuations*. Each bonus valuations is parameterized by four parameters: $t, k, P(\cdot), B(\cdot)$, where $t, k$ are non-negative numbers, $B : 2^M \to \{0,1\}$ is a boolean function and $P : 2^M \to \{0,1\}$ is a monotone boolean function, both described by polynomial-size circuits. Hence, the complete representation $(t, k, P, B)$ of a bonus valuations takes polynomial space. Given these parameters, the following valuation $v$ is a bonus valuation:

$$v(S) = \begin{cases} |S| \cdot t, & \text{if } |S| < k; \\ (k - \frac{1}{2^{|S|}}) \cdot t, & \text{if } |S| \geq k \text{ and } P(S) = 0 \\ k \cdot t - \frac{1}{m^4}, & \text{if } |S| = k, P(S) = 1 \text{ and } B(S) = 0; \\ k \cdot t, & \text{if } |S| = k, P(S) = 1 \text{ and } B(S) = 1; \\ k \cdot t, & \text{if } |S| > k \text{ and } P(S) = 1. \end{cases}$$

Notice that computing the value $v(S)$ for a given bundle $S$ (that is, implementing value queries) for bonus valuations can be done in polynomial time, given the representation above. Also, similar to [4], it can be verified that $v(S)$ is a monotone submodular function for any boolean function $B : 2^M \to \{0, 1\}$ and monotone $P : 2^M \to \{0, 1\}$.

The definition of the bonus function is similar but not identical to the one in [4]. To assist the reader who is familiar with the proof in [4], here is how the specific functions $P(\cdot)$ and $B(\cdot)$ that we choose later will roughly correspond to the construction in [4]: For sets of size $k$, the function $P(S)$ describes the "structured menu", by giving value 1 to bundles in the structured menu and 0 otherwise. It also specifies the bundles that are larger and more expensive than the bundles in the structured menu, by giving value 1 to them. Among the bundles in the structured menu, the function $B(\cdot)$ picks the bonus bundles by giving value 1.

## 2.2 The Reduction

Our goal is to solve a SAT instance, using the presumed truthful mechanism. Recall the following concept from [4]: $\mathcal{R}_{v_{-i}}$ is the menu induced by $v_{-i}$, i.e. the collection of all possible sets that could be allocated to bidder $i$, assuming that the $(n-1)$-tuple of valuations of other bidders is $v_{-i}$. If a set $S \in \mathcal{R}_{v_{-i}}$ is allocated to bidder $i$, he is charged a price $p_{v_{-i}}(S)$ that does *not* depend on his own declared valuation (the "taxation principle"). Some sets are not on the menu, but for convenience we extend the definition of $p_{v_{-i}}(S)$ to $S \notin \mathcal{R}_{v_{-i}}$, by setting $p_{v_{-i}}(S) := \min_{T \in \mathcal{R}_{v_{-i}} : S \subset T} p_{v_{-i}}(T)$. If there no superset of $S$ on the menu, we define $p_{v_{-i}}(S) = \infty$.

**Definition 2.2** (Structured Submenu). *Given valuations $v_{-i}$ of all bidders $i' \neq i$, and parameters $k, p \in [0, m]$, the structured submenu $\mathcal{S}(v_{-i}, k, p)$ is the collection of all sets if items $S$ such that*

- $S \in \mathcal{R}_{v_{-i}}$,
- $|S| = k$,
- $p - \frac{1}{m^5} < p_{v_{-i}}(S) \leq p$,
- *for all $T \in \mathcal{R}_{v_{-i}}$ such that $T$ strictly contains $S$, $p_{v_{-i}}(T) - p_{v_{-i}}(S) \geq \frac{1}{m^3}$.*

*We call a bundle $S$ a* candidate, *if it satisfies the conditions above (i.e. $S \in \mathcal{S}(v_{-i}, k, p)$, when $\mathcal{S}(v_{-i}, k, p)$ will be clear from the context).*

**Lemma 2.3** (essentially from [4]). *Let $A$ be a (deterministic) truthful $\frac{n}{10}$-approximation mechanism for combinatorial auctions with submodular valuations. Let $v_{-i}$ be a set of valuations where each valuation in $v_{-i}$ is a random polar additive valuation (with $m/n$ items of value 1). Then, with a constant probability, there exists $k \in \{1, \ldots, m\}$ and $p \in [0, m]$ (a multiple of $1/m^5$) such that $|\mathcal{S}(v_{-i}, k, p)| \geq \frac{e^{\frac{m}{n^2}}}{10n^2 \cdot m^6}$.*

In [4] only the existence of such a structured menu was proven. However, it is easy to see that the proof of [4] actually shows that this event occurs with constant probability as well. Specifically, the proof works as follows: we select at random a player $i$ and construct random polar additive valuations for all other players. Then, with constant probability there are parameters $p$ and $k$ such that the respective structured submenu is sufficiently large.

In our construction we need to find $k$ and $p$ in polynomial time. We do that as follows. Observing that $k$ can be one of $m$ possible values $1, \ldots, m$ and that $p$ can be one of $m^6$ values $\frac{1}{m^5}, \frac{2}{m^5}, \ldots, m$, we notice that number of possible combinations of values for $p$ and $k$ is polynomially bounded. Therefore we can try each combination one by one. The rest of the analysis assumes that the values of $k, p$ and the random coins are successful in the sense that the structured submenu $\mathcal{S}(v_{-i}, p, k)$ has cardinality at least $\frac{e^{\frac{m}{n^2}}}{10n^2 \cdot m^6}$.

Next, we set up a bonus valuation for bidder $i$ that will allow us to embed the SAT problem in it. We choose the parameter $t$ to be equal to $t = 2^{2m}$. The parameters $k$ and $p$ of the bonus valuation are identical to the parameters $k$ and $p$ of the structured submenu. The following lemma describes the function $P_p(\cdot)$ that we use as the $P(\cdot)$ parameter of the bonus valuation.

**Lemma 2.4.** *Let $A$ be a truthful mechanism for combinatorial auctions with submodular bidders. Fix valuations $v_{-i}$ for all bidders except $i$, and parameters $k \in \mathbb{Z}, p \in [0, m]$. Let $P_p(S)$ be the following function:*

$$P_p(S) = \begin{cases} 1 & \text{if } |S| = k \text{ and } S \text{ is a candidate } (S \in \mathcal{S}(v_{-i}, k, p)); \\ 1 & \text{if } |S| > k \text{ and } p_{v_{-i}}(S) > p; \\ 0 & \text{otherwise.} \end{cases}$$

*Then $P_p(\cdot)$ is a monotone boolean function and $P_p(S)$ can be evaluated in time polynomial in $n$, $m$, and the running time of $A$.*

*Proof.* Let us verify that $P_p(\cdot)$ is a monotone function. Let $S \subset T$. If $P_p(S) = 1$ then one possibility is that $S$ is a candidate and thus by definition of a structured menu, $p_{v_{-i}}(T) > p$. The other possibility is that $|S| > k$ and $p_{v_{-i}}(S) \geq p$. Then we also have $|T| \geq |S| > k$ and $p_{v_{-i}}(T) \geq p_{v_{-i}}(S) > p$ by the monotonicity of the payment function $p_{v_{-i}}$. In both cases, $P_p(T) = 1$.

To evaluate $P_p(S)$, we show that we can efficiently decide whether $S$ is a candidate and whether $p_{v_{-i}}(S) > p$ or not, using the truthful mechanism as a black box. Given a bundle $S$, consider the following additive valuation $v$ for bidder $i$: $v(\{j\}) = 2p$ if $j \in S$ and $v(\{j\}) = 0$ otherwise. Let $S'$ be the bundle allocated to $i$ in $A(v, v_{-i})$ and let $p'$ be the price that $i$ is charged in this instance.

First, assume that $|S| = k$; here we want to determine whether $S$ is a candidate. We start with showing that if $S$ is a candidate, then the returned bundle under valuation $v$ must be $S' = S$. To see this, recall that if $S$ is a candidate, then $S \in \mathcal{R}_{v_{-i}}$ and $p_{v_{-i}}(S) \leq p$, therefore the profit from $S$ under valuation $v$ is $v(S) - p_{v_{-i}}(S) \geq 2p|S| - p$. No other set can have higher profit, because supersets of $S$ have the same value and strictly higher price, while sets that do not contain $S$ have value at most $2p|S| - 2p$. Thus, $S' = S$ and we also learn the price $p_{v_{-i}}(S)$. If $p_{v_{-i}}(S) \in [p - \frac{1}{m^4}, p]$, it remains to decide whether for each $T$ that strictly contains $S$, $p_{v_{-i}}(S) + \frac{1}{m^3} \leq p_{v_{-i}}(T)$. This can be done by altering $v$ to another additive valuation $v_j$ which is identical to $v$ except that $v_j(\{j\}) = \frac{1}{m^3}$, for exactly one item $j$. Observe that $p_{v_{-i}}(S) + \frac{1}{m^3} \leq p_{v_{-i}}(T)$ for all $T \supsetneq S$ if and only if $i$ is allocated $S$ in $A(v_{-i}, v_j)$ for every $j \notin S$. In this case, we conclude that $S$ is a candidate. If the above process fails at at point, $S$ cannot be a candidate.

Next, we consider the case where $|S| > k$; here we just need to check whether $p_{v_{-i}}(S) > p$. If $S' = S$ then we learn the price of $S$ and we can immediately check whether $p_{v_{-i}}(S) > p$. Next, consider the case where there exists some item $j \in S \setminus S'$. Then we claim that $p_{v_{-i}}(S) > p$. To see this, observe that $p_{v_{-i}}(S) - p' \geq 2p$ since the marginal value of $j$ is $2p$. Since $p'$ is nonnegative we have that $p_{v_{-i}}(S) \geq 2p > p$.

The last remaining case is that $S \subset S'$. Then we claim that $p_{v_{-i}}(S) = p_{v_{-i}}(S')$: The price function $p_{v_{-i}}(\cdot)$ is monotone so we have that $p_{v_{-i}}(S) \leq p_{v_{-i}}(S')$. Moreover, our construction of $v$ implies that $v(S') = v(S)$ and thus if $p_{v_{-i}}(S') > p_{v_{-i}}(S)$ then the profit from $S$ is larger than the profit from $S'$. But this cannot be since $S$ is allocated to $i$ by the (truthful) mechanism. So we can again check whether $p_{v_{-i}}(S) = p_{v_{-i}}(S') > p$. □

We now proceed to implementing the function $B(\cdot)$. The role of this function is to specify which candidate bundles get a bonus. A naive way to do so would be to explicitly list the bundles that get a bonus. However, this would make finding a profit-maximizing bundle too easy. Thus, we use an implicit way of specifying these bundles: all bundles that correspond to a satisfying assignment of a certain SAT instance. Specifically, given a boolean formula $\phi$ on $\ell$ variables, we construct a circuit for the following function $B(\cdot)$. We generate a uniformly random matrix $T \in \{0,1\}^{\ell \times m}$, which induces a linear mapping $T : \{0,1\}^m \to \{0,1\}^\ell$ where $T(S) = T \cdot \mathbf{1}_S$, all operations modulo 2. We let $B(S) = 1$ if and only if the assignment of the $\ell$ variables defined by $T(S)$ satisfies $\phi$.

**Claim 2.5.** *Suppose that $\phi$ is satisfiable and $x \in \{0,1\}^\ell$ is a satisfying assignment. If $\mathcal{S} \subseteq \{0,1\}^m$, $|\mathcal{S}| > 2^{2\ell}$, then with probability at least $1 - 2^{-\ell}$ there exists some $S \in \mathcal{S}$ such that $T(S) = x$.*

*Proof.* Let $Z_S$ denote the indicator variable of the event $T(S) = x$, and let $Z = \sum_{S \in \mathcal{S}} Z_S$. We can assume that $\emptyset \notin \mathcal{S}$; removing the empty set decreases $|\mathcal{S}|$ only by 1. Then, for any $S \in \mathcal{S}$, we have $\mathbf{E}[Z_S] = \Pr[T(S) = x] = 2^{-\ell}$, since $T(S)$ is distributed uniformly in $\{0,1\}^\ell$. Therefore, $\mathbf{E}[Z] = 2^{-\ell}|\mathcal{S}|$.

Moreover, for any $S \neq S' \in \mathcal{S}$, we have $\mathbf{E}[Z_S Z_{S'}] = \Pr[T(S) = T(S') = x] = 2^{-2\ell}$. This is because $T(S') = T(S \cap S') \oplus \sum_{i \in S' \setminus S} T_i$ where $T_i$ is the $i$-th column of $T$, uniformly distributed in $\{0,1\}^\ell$ and independent of $T(S)$ for $i \notin S$. (We can assume WLOG that $S' \setminus S \neq \emptyset$.) Therefore even conditioning on $T(S) = x$, the probability that $T(S') = x$ is still $2^{-\ell}$.

In other words, the variables $Z_S$ are pairwise independent and we can compute $\mathbf{Var}[Z] = \sum_{S \in \mathcal{S}} \mathbf{Var}[Z_S] = \sum_{S \in \mathcal{S}} (\mathbf{E}[Z_S^2] - \mathbf{E}[Z_S]^2) = (2^{-\ell} - 2^{-2\ell})|\mathcal{S}|$. By Chebyshev's inequality,

$$\Pr[Z = 0] \leq \frac{\mathbf{Var}[Z]}{(\mathbf{E}[Z])^2} = \frac{(2^{-\ell} - 2^{-2\ell})|\mathcal{S}|}{(2^{-\ell}|\mathcal{S}|)^2} \leq \frac{1}{2^{-\ell}|\mathcal{S}|} \leq \frac{1}{2^\ell}.$$

□

Finally, we prove that if $\phi$ is satisfiable and all the random choices are successful in the sense that the structured submenu $\mathcal{S}(v_{-i}, k, p)$ is sufficiently large and the random transformation $T$ hits the satisfying assignment, then the mechanism must return a bundle $S'$ such that $B(S') = 1$.

**Lemma 2.6.** *If bidder $i$'s valuation $v$ is a bonus valuation as constructed above and there exists $S^* \in \mathcal{S}(v_{-i}, k, p)$ such that $T(S^*)$ is a satisfying assignment to the formula $\phi$, then bidder $i$ receives a bundle $S'$ such that $B(S') = 1$.*

*Proof.* The proof is very similar to one that can be found in [4] and follows from the following series of claims. We fix some bundle $S^* \in \mathcal{S}(v_{-i}, k, p)$ such that $B(S^*) = 1$.

**Claim 2.7.** $v(S^*) - p_{v_{-i}}(S^*) > v(S) - p_{v_{-i}}(S)$, *for every $S$ such that $|S| < k$ or $P(S) = 0$.*

*Proof.* We have $v(S^*) - p_{v_{-i}}(S^*) \geq kt - p \geq kt - m$. If $|S| < k$, then clearly $v(S^*) - p_{v_{-i}}(S^*) > v(S) - p_{v_{-i}}(S)$, because $v(S) \leq (k-1)t < kt - m$ (since $t = 2^{2m}$). If $|S| \geq k$ and $P(S) = 0$, then $v(S) = (k - 1/2^{|S|})t \leq (k - 1/2^m)t < kt - m$, again because $t = 2^{2m}$. □

**Claim 2.8.** $v(S^*) - p_{v_{-i}}(S^*) > v(S) - p_{v_{-i}}(S)$, *for every $S$ such that $|S| > k$ and $P(S) = 1$.*

*Proof.* Here, $v(S) \leq kt = v(S^*)$. To finish the proof, observe that by definition of $P(S)$ we have $p_{v_{-i}}(S) > p \geq p_{v_{-i}}(S^*)$. □

**Claim 2.9.** $v_i^{S^*}(S^*) - p_{v_{-i}}(S^*) > v_i^{S^*}(S) - p_{v_{-i}}(S)$, *for every $S$ such that $|S| = k$, $P(S) = 1$ and $B(S) = 0$.*

*Proof.* Hence $v(S) = kt - \frac{1}{m^4}$, while $v(S^*) = kt$. By the properties of the structured submenu, $|p_{v_{-i}}(S) - p_{v_{-i}}(S^*)| \leq \frac{1}{m^5}$ which proves the claim. □

Therefore, the only possible bundles allocated to bidder $i$ have $|S| = k$, $P(S) = 1$ and $B(S) = 1$. □

Now we can finish and summarize the proof of Theorem 2.1.

*Proof.* Let $\epsilon > 0$ be any positive constant. We assume that there a truthful mechanism for combinatorial auctions with $n$ submodular bidders and $m$ items that achieves a $\frac{n}{10}$-approximation, when $n = 10m^{1/2-\epsilon}$. Our goal is to solve a SAT problem in which we are given a formula $\phi$ with $\ell$ variables.

We choose parameters $n = 10m^{1/2-\epsilon} = poly(\ell)$ so that $\frac{e^{\frac{m}{n^2}}}{10n^2 \cdot m^6} > 2^{2\ell}$, and we produce an instance of combinatorial auctions with $n$ bidders and $m$ items. We set $t = 2^{2m}$. We pick a random bidder $i$ and random polar additive valuations $v_{-i}$ for bidders $i' \neq i$. Then we try all possible values of $k \in \{1, 2, \ldots, m\}$ and $p \in \{\frac{1}{m^5}, \frac{2}{m^5}, \ldots, m\}$ and we construct a bonus valuation for bidder $i$ with these parameters. We construct a circuit implementing the function $P(\cdot)$, by simulating the computation of the presumed mechanism, according to Lemma 2.4. We also pick a random matrix $T \in \{0,1\}^{\ell \times m}$ and construct a circuit implementing the function $B(\cdot)$ such that $B(S) = 1$ if and only if $T(S)$ satisfies $\phi$. Then we run the mechanism on the bonus valuation defined by $(t, k, P, B)$ for bidder $i$ and $v_{-i}$ for the remaining bidders.

If the mechanism returns a bundle $S$ for bidder $i$ such that $B(S) = 1$, we have found a satisfying assignment $T(S)$ to the formula $\phi$. If not, we repeat this process polynomially many times. If the process never succeeds, we answer that the formula is unsatisfiable. This procedure can fail only if the formula is satisfiable, with exponentially small probability. □

## 3 Truthful-in-Expectation Mechanisms for Combinatorial Auctions

In this section, we want to rule out the existence of any constant-factor *truthful-in-expectation* mechanism, as opposed to deterministic and universally truthful mechanisms. In fact, the proof holds for the stronger variant of approximately truthful-in-expectation mechanisms.

**Definition 3.1.** *For $\epsilon \geq 0$, a mechanism with allocation and payment rules $\mathcal{A}$ and $p$ is $(1-\epsilon)$-approximately truthful-in-expectation if every player $(1-\epsilon)$-approximately maximizes his expected utility by truthfully reporting his valuation function, meaning that*

$$\mathbf{E}[v_i(\mathcal{A}(v)) - p_i(v)] \geq (1-\epsilon)\mathbf{E}[v_i(\mathcal{A}(v'_i, v_{-i})) - p_i(v'_i, v_{-i})]$$

*for every player $i$, (true) valuation function $v_i$, (reported) valuation function $v'_i$, and (reported) valuation functions $v_{-i}$ of the other players. The expectations above are over the coin flips of the mechanism.*

We prove the following.

**Theorem 3.2.** *There are absolute constants $\epsilon, \gamma > 0$ such that for every fixed $n$, there is a class of succinctly represented monotone submodular valuations, for which there is no $n^{-\gamma}$-approximation truthful-in-expectation (even $(1-\epsilon)$-approximately truthful-in-expectation) mechanism for combinatorial auctions with $n$ players, unless $NP \subset P/poly$.*

The proof of this theorem builds on several recent advances and techniques that have been developed in this area. Therefore, before we describe the proof in detail, we summarize its main components.

**Proof background** We start from the hardness proof in the value oracle model that was developed in [12]. The proof of [12] is based on a *density-boosting technique* which implies that every mechanism for combinatorial auctions with submodular bidders must either use an exponential number of value queries, or we obtain a contradiction with the property of truthfulness in expectation. The contradiction is obtained from a technical argument relying on the *symmetry gap* machinery of [22]. Recently, [9] proposed a generic way of converting hardness proofs using this machinery into *computational hardness* proofs, for succinctly represented submodular functions. This reduction uses list decodable codes and the known hardness of Unique-SAT (deciding whether a formula has 0 or 1 satisfying assignments). We appeal to the same idea here, and represent submodular valuation functions in the same format as [9].

A price that we have to pay for using the framework of [9] is that the symmetry gap argument can be used only with parameters that lead to hardness depending on the number of players, but not the number of items ($n^{-\gamma}$, as opposed to $m^{-\gamma}$). Possibly these parameters can be optimized to yield hardness with a (mild) dependence on $m$ (depending on the best currently known list-decodable codes), but for ease of exposition we restrict ourselves to ruling out $n^{-\gamma}$-approximation in this paper.

Following the framework of [9], assuming that there exists an approximately-TIE mechanism providing a good approximation for these succinctly represented functions, we provide a way of solving Unique-SAT using the mechanism. The way we embed the Unique-SAT problem in an instance of combinatorial auctions is based on the density-boosting technique of [12]. Unfortunately, the parameters describing the embedding are algorithmically inaccessible and must be provided as an advice string. The last technical question we have to deal with is whether these parameters can be described by a polynomially-bounded string for each input size. This requires some additional work and modification of certain technical lemmas from [12]. In conclusion, we show how to use the presumed mechanism to solve Unique-SAT using polynomial advice, and this implies $NP \subset RP/poly = P/poly$.

**Proof sketch** Now we describe roughly how the proof works. We use repeatedly the *taxation principle*, which in the case of truthful-in-expectation mechanisms says the following: For each player $i$, once the valuations of the other players are fixed, the mechanism implicitly offers a *menu* $\mathcal{M}$ whose entries are distributions over sets $S_i$ and prices $p_i$. The menu consists of all the possible distributions that the player could receive when reporting a certain valuation. The taxation principle (which follows directly from the property of truthfulness) states that for each valuation $v_i$ reported by player $i$, the mechanism must allocate to him the distribution from the menu that maximizes the expected utility $\mathbf{E}[v_i(S_i) - p_i]$.

We start from a basic instance, as in [4], where each player $i$ has a random desired set $A_i$. As was proved in [4], the approximation guarantee implies that there is a player $i$ whose allocation in this instance is at least somewhat correlated with his desired set (to an extent that depends on the approximation factor). In the following, we fix this "special player" as well as the valuations of the other players.

Next, we consider more complicated valuations for the special player, that we call "double-peak valuations". These valuations were developed in [12]. Each double-peak valuation is supported on the union of two sets $A \cup B$ and is defined in such a way that the sets $A, B$ have a "high value", while sets that are evenly split by the partition $(A, B)$ have a lower value. The technical obstacle here is that we want to present the valuations *succinctly* on the input, without explicitly revealing the partition $(A, B)$. To achieve this, we use the list-decoding idea of [9].

The reason why double-peak valuations are useful in proving a hardness result is quite technical. One indication of why these functions might be useful is that the analytic transformation that is crucial in the truthful-in-expectation mechanism for coverage functions in [11] does not work for double-peak valuation functions. More precisely, in contrast to coverage valuations where this transformation produces a *concave function*, for a double-peak valuation the resulting function is not concave. This is the starting point of the hardness proof in the oracle model in [12]. Here, we adapt the machinery of [12] to our purpose as follows.

As in [12], we consider double-peak valuations at different "levels", where the cardinality of the support $A \cup B$ doubles at each level. We measure the "value at level $j$" of distributions on the menu of the special player by the double-peak valuations supported on $A \cup B$, where $A, B$ are chosen randomly at level $j$. We prove that there must be two successive levels $j, j+1$ where there is a relatively valuable distribution on the menu with respect to level $j$, but no distribution is very valuable with respect to level $j+1$. This follows from the density-boosting technique of [12]: If this were not the case, we would obtain by an inductive argument a very valuable distribution at the last level, which would lead to a contradiction with the taxation principle.

An added difficulty here is that the notion of being valuable also incorporates prices, by way of a multiplier $\lambda$ that serves as a conversion factor between values and prices. We have to ensure that the necessary multiplier $\lambda$ can be described by polynomially many bits. This leads to some technical issues that we deal with in this paper. In conclusion, we have a polynomial-size description of the "place to embed" a valuation, with respect to which a valuable distribution exists on the menu, but it cannot be found if we do not know the partition $(A, B)$.

In fact, the way we measure value at level $j+1$ corresponds to what the mechanism can achieve at level $j$, if the output distribution is "balanced with respect to the partition $(A, B)$". Since we make it a computationally difficult task to find the partition $(A, B)$, we would in fact expect the output distribution to be balanced with respect to $(A, B)$. But we prove that there exists a distribution on the menu, strictly more valuable than any that is balanced with respect to $(A, B)$. So truthfulness in expectation implies that the mechanism must in fact return an *unbalanced distribution*.

Finally, we show how to solve the Unique-SAT problem, assuming the scenario above. Using the list-decoding idea of [9], we set up a submodular valuation encoded by a Unique-SAT formula at level $j$, with the property that if the Unique-SAT formula is satisfiable, that the player should receive a relatively valuable distribution from the mechanism. The satisfying assignment corresponds to the unknown partition $(A, B)$ which is not easily determined by the mechanism, given the succinct representation of the valuation. The parameters necessary for setting up the submodular valuation cannot be all computed algorithmically, but we ensure that for each input size $n$ they can be described by $poly(n)$ bits. These parameters will be given as an advice string in the reduction. Finally, the property that there is no very valuable balanced distribution on the menu guarantees that the mechanism must return an unbalanced distribution, in case the formula is satisfiable. This would allow us to determine the partition $(A, B)$ and find a satisfying assignment. Hence, we are able to decide the Unique-SAT problem using polynomial advice, which implies that $NP \subset P/poly$.

## 3.1 The Bidding Language

Here, we define two classes of submodular functions along with the encoding that will be used in the hardness proof.

**Polar additive valuations** The first type of function that we use (following [4]) is simple:

- $v_{A,\omega}(S) = |S \cap A| + \omega |S \setminus A|$, where $A$ is a set and $\omega > 0$ is a positive real number.

Such functions are monotone, additive and hence also submodular. We encode them naturally by specifying the set $A$ and the parameter $\omega$.

**Double-peak valuations** The second type of function (following [12]) is one we call "double-peak valuation", because each has two disjoint sets $A, B$ of "high value". The function is parameterized by disjoint sets $A, B$ of equal cardinality and real numbers $\alpha, \beta > 0$. The value of $f_{A,B,\alpha,\beta}(S)$ is defined as follows:

- Let $x = \frac{|S \cap A|}{|A|}$ and $y = \frac{|S \cap B|}{|B|}$. Then $f_{A,B,\alpha,\beta}(S) = \tilde{\psi}(x, y)$ where $\tilde{\psi}(x, y)$ is defined as follows:[2]

- If $|x - y| \leq \beta$, then $\tilde{\psi}(x, y) = 1 - (1 - \frac{x+y}{2\alpha})_+^2$

- If $x - y > \beta$, then $\tilde{\psi}(x, y) = 1 - (1 - \frac{2x-\beta}{2\alpha})_+ (1 - \frac{2y+\beta}{2\alpha})_+$

- If $y - x > \beta$, then $\tilde{\psi}(x, y) = 1 - (1 - \frac{2x+\beta}{2\alpha})_+ (1 - \frac{2y-\beta}{2\alpha})_+$

We also define a "symmetrized double-peak function" $\bar{f}_{A \cup B, \alpha}$ by

- $\bar{f}_{A \cup B, \alpha}(S) = \tilde{\psi}(\frac{x+y}{2}, \frac{x+y}{2}) = 1 - (1 - \frac{x+y}{2\alpha})_+^2$ where $x, y$ are as above.

Note that $\bar{f}_{A \cup B, \alpha}(S)$ does not depend on $\beta$ and the partition of $A \cup B$ into $A, B$, which justifies our notation.

It is shown in [12] that these functions are monotone submodular. They could be encoded by specifying the quadruple $(A, B, \alpha, \beta)$, or the pair $(A \cup B, \alpha)$ in the symmetric case, but that is not the encoding we use. Instead, we apply a more intricate encoding using list decodable codes, developed in [9]. We warn the reader that this encoding may not look very natural. All we ask is that an encoding $\Upsilon$ defines a submodular function $v_\Upsilon$ such that given $\Upsilon$ and a set $S$, the value of $v_\Upsilon(S)$ can be computed efficiently. In particular, this encoding allows one to use any algorithm in the value oracle model, such as the (non-truthful) $(1 - 1/e)$-approximation for welfare maximization in combinatorial auctions [21]. In contrast, we prove that the requirement of truthfulness in expectation prevents one from achieving any constant-factor approximation for the same submodular valuations.

---

[2] We define $(x)_+ = \max\{x, 0\}$ to be the positive part of $x$. By $(x)_+^2$, we mean $(\max\{x, 0\})^2$.

We will need the following definition.

**Definition 3.3** (List Decodable Codes). *A pair of functions $(E, D)$, $E : \Sigma^m \to \Sigma^n$, $D : \Sigma^n \to (\Sigma^m)^\ell$, $\ell = poly(n)$ is called an $(n, m, d)$-list decodable code if:*

1. *$E$ is injective.*

2. *For $c \in \Sigma^n$ and $x \in \Sigma^m$, $x \in D(c)$ if and only if $d_H(c, E(x)) \leq d$.*

Here, $d_H$ denotes Hamming distance, $d_H(x, y) = |\{i : x_i \neq y_i\}|$. In this paper we are only interested in cases where $E$ and $D$ can be computed in polynomial time.

In particular, we use list decodable codes over $\Sigma = \{0, 1\}$, i.e. binary codes. We fix a family of binary $(m'', m', \frac{1}{2}(1 - \beta)m'')$-list decodable codes for all sufficiently large $m'' \in \mathbb{Z}_+$ (over $\Sigma = \{0, 1\}$, with $m'' = poly(m')$ and constant $\beta > 0$; such codes are described for instance in [15]). We denote the encoding function by $E : \{0, 1\}^{m'} \to \{0, 1\}^{m''}$ and the decoding function by $D : \{0, 1\}^{m''} \to (\{0, 1\}^{m'})^\ell$. These codes are fixed in our reduction and assumed to be implicitly described by the parameters $m'$, $m''$ and $\beta$.

**The encoding of double-peak valuations** We encode a valuation function by parameters $\alpha, \beta > 0$, an *ordered set* $C$ of cardinality $|C| = 2m''$ and a *boolean formula* $\phi$ on $m'$ variables, which is assumed to have at most one satisfying assignment. (If $\phi$ has multiple satisfying assignments, it does not encode a submodular function and we do not assume anything about how the mechanism behaves on such input.) Given $(C, \phi, \alpha, \beta)$, we define $v_{C, \phi, \alpha, \beta}(S)$ as follows.

- If $\phi$ does not have any satisfying assignment, then we define $v_{C, \phi, \alpha, \beta}(S) = \bar{f}_{C, \alpha}(S)$ (see the definition of $\bar{f}_{A \cup B, \alpha}$ above, where we set $A \cup B = C$, ignoring the ordering of the elements of $C$).

- If $\phi$ has a (unique) satisfying assignment $x = (x_1, \ldots, x_{m'})$, let $y = E(x) \in \{0, 1\}^{m''}$. We identify the ordered set $C$ in some canonical way with $[m''] \times \{0, 1\}$ (for example by a bijection which maps the ordering of $C$ to the lexicographic ordering of $[m''] \times \{0, 1\}$) and we define $A$ by an expansion procedure $A = exp(y)$ as follows: If $y_i = 0$, then $(i, 0) \in A$ and $(i, 1) \notin A$; if $y_i = 1$, then $(i, 0) \notin A$ and $(i, 1) \in A$. We also define $B = C \setminus A$. Note that $|A| = |B| = m''$. We define $v_{C, \phi, \alpha, \beta}(S) = f_{A, B, \alpha, \beta}(S)$.

We show that $(C, \phi, \alpha, \beta)$ is indeed a legitimate representation of $v_{C, \phi, \alpha, \beta}$ in the sense that the function can be evaluated efficiently, given $(C, \phi, \alpha, \beta)$.

**Lemma 3.4.** *Given $(C, \phi, \alpha, \beta)$, the value of $v_{C, \phi, \alpha, \beta}(S)$ can be calculated in polynomial time for any $S$.*

*Proof.* First, the value of the function depends only on the elements in $C$, so we might as well restrict the ground set to $C$ for now. We view subsets of $C$ as strings in $\{0, 1\}^{[m''] \times \{0, 1\}}$; the Hamming distance $d_H$ is then equivalent to the symmetric difference between sets. Observe that if $S$ is *balanced* in the sense that $||S \cap A| - |S \cap B|| \leq \beta m''$, we can calculate the value of $v_{C, \phi, \alpha, \beta}(S)$ without having to know the partition $C = A \cup B$, because we are in the symmetric case where the value depends only on $|S \cap C|$. Therefore, we have to show a polynomial-time procedure that finds out whether $S$ is balanced or unbalanced and finds the partition $(A, B)$ in the unbalanced case. In the following we show a procedure that identifies whether $|S \cap A| - |S \setminus B| > \beta m''$ and in that case finds $(A, B)$. The other case is similar (by considering the complement of $S$ in $C$).

**Claim 3.5.** *Let $S \subseteq A \cup B$, $A \cap B = \emptyset$ and $|A| = |B| = m''$. If $|S \cap A| - |S \setminus B| > \beta m''$ then $d_H(S, A) < (1 - \beta)m''$.*

*Proof.* Using the facts that $S \subseteq A \cup B$, $A \cap B = \emptyset$ and $|A| = m''$,

$$\beta m'' < |S \cap A| - |S \cap B| = m'' - |A \setminus S| - |S \setminus A| = m'' - d_H(S, A).$$

□

For $a \in \{0,1\}$, we define a contracting procedure $con^a : 2^{[m''] \times \{0,1\}} \to \{0,1\}^{m''}$ that takes a set $S \subseteq C \simeq [m''] \times \{0,1\}$ and produces the following $m''$-bit string $y = con^a(S)$: If $(i,0) \notin S$ and $(i,1) \in S$ then $y_i = 0$. If $(i,0) \in S$ and $(i,1) \notin S$ then $y_i = 1$. Otherwise, $y_i = a$. Note that considering the expansion procedure above, $con^a(exp(y)) = y$ for any $y \in \{0,1\}^{m''}$ and $a \in \{0,1\}$. However, some subsets of $C$ are not the expansion image $exp(y)$ of any $y \in \{0,1\}^{m''}$. We now need the following simple claim:

**Claim 3.6.** *If $A = exp(y)$ and $d_H(S, A) < (1-\beta)m''$ then there exists $a \in \{0,1\}$ such that $d_H(con^a(S), con^a(A)) < (1-\beta)\frac{m''}{2}$.*

*Proof.* Consider a pair of elements $(i,0), (i,1)$. Since we assume $A = exp(y)$, $A$ contains exactly one of the two elements. If $S \cap \{(i,0),(i,1)\} = A \cap \{(i,0),(i,1)\}$, then $con^a(S)$ and $con^a(A)$ will agree in the $i$-th bit, regardless of $a$. If $S$ contains one of these elements and not the same one as $A$, then $con^a(S)$ and $con^a(A)$ will disagree in the $i$-the bit, regardless of $a$. If $S$ contains both or neither element, then $con^a(S)$ and $con^a(A)$ agree in the $i$-bit for one of the values $a = 0$ or $a = 1$, and not for the other. In summary, two disagreements between $S$ and $A$ translate to one bit of disagreement after applying $con^a$, and one disagreement between $S$ and $A$ translates to a bit with $\frac{1}{2}$ probability of disagreement (in expectation over choosing $a \in \{0,1\}$ at random). Thus the claim must hold for some value of $a$. □

To complete the proof, we observe that if $x$ is a satisfying assignment to the formula $\phi$, $A = exp(E(x))$ and $d_H(con^a(S), E(x)) = d_H(con^a(S), con^a(A)) < (1-\beta)\frac{m''}{2}$ then the list-decoding property of the code implies that the unique satisfying assignment $x$ must be one of the polynomially many strings that $D(con^a(S))$ returns. We can check if $x$ is one of the strings of $D(con^a(S))$ simply by testing the satisfiability of each of the assignments that $D(con^a(S))$ returns, for each of the two possible values of $a$. If none of the assignments is satisfying, we know that $S$ is balanced with respect to $(A, B)$ and we do not need the knowledge of the partition $(A, B)$ to compute $v_{C,\phi,\alpha,\beta}(S) = \bar{f}_{C,\alpha}(S)$. If we find the partition $(A, B)$, computing $v_{C,\phi,\alpha,\beta}(S) = f_{A,B,\alpha,\beta}(S)$ is straightforward. □

The formal proof of computational hardness for combinatorial auctions with these valuations can be found in Appendix A.

## 4 Truthful-in-Expectation Mechanisms for Exact Coverage CPP

We now turn our attention to combinatorial public projects with coverage valuation. A coverage valuation is defined as follows: there exists a universe $U$, and each item of the auction $j \in M$ is identified with a subset $S_j \subseteq U$. Now we define the valuation $v$ as $v(S) = \cup_{j \in S} S_j$.

**Theorem 4.1.** *Unless $NP \subseteq P/poly$, there is no constant-factor truthful-in-expectation approximation for the exact CPP problem with coverage valuations, even for 1 bidder (i.e. the problem $\max\{f(S) : |S| = k\}$ where $f$ is a coverage function).*

Let us point out some aspects of this result and put it in context. For (exact) CPP a significant gap in approximability between truthful vs. non-truthful mechanisms was proved in [20]. In particular, this separation was proved for deterministic mechanisms and the class of submodular functions. The proof followed the characterize and optimize approach, involving a characterization of all deterministic truthful mechanisms and then a bound of VCG-based mechanisms by an application of the Sauer-Shelah lemma. The proof was subsequently simplified in [2], where it was proved that no deterministic truthful mechanism can achieve a $m^{1/2-\epsilon}$-approximation for coverage functions (a subclass of submodular functions), still relying the Sauer-Shelah lemma.

Our result rules out not only deterministic but also randomized *truthful-in-expectation* mechanisms. While the previous proofs can be extended to universally truthful mechanisms quite easily, truthful-in-expectation mechanisms maybe more powerful in this setting. In particular, [10] proved that the *flexible* CPP problem, where a mechanism is allowed to return a set $S$ of size at most $k$, admits a truthful-in-expectation $(1 - 1/e)$-approximation for coverage functions. Therefore, we are right on the boundary between tractable and intractable problems, in terms of truthfulness in expectation. The first hardness for truthful-in-expectation mechanisms has been obtained only recently, in the value oracle model (for the exact CPP problem in [4], for the flexible CPP problem and combinatorial auctions in [12]). Here we prove a *computational hardness* result, for the natural class of coverage valuations.

*Proof overview.* Our proof is very simple, directly from the hardness of Max $k$-cover, using an argument somewhat similar to [12]. In particular, we do not use the Sauer-Shelah lemma or any similar combinatorial tool. The proof follows an outline somewhat similar to [12]; it is easier to explain if we assume that the mechanism is maximal-in-distributional-range (MIDR), or in other words that all prices are equal to 0. Then, the mechanism must maximize over a certain range of distributions $\mathcal{R}$. Suppose the mechanism provides a $c$-approximation. First, we prove that for every set $A$ of size $k$, the range $\mathcal{R}$ must contain a distribution that takes at least a $c$-fraction of $A$ in expectation - this follows directly from the property of $c$-approximation. Therefore, if we run $\mathcal{M}$ on an instance of Max $k$-cover that that has a hidden optimal solution of size $k$, the range contains a solution that is somewhat correlated with the optimal solution. Therefore, optimizing over the range will reveal whether the optimum is "high" or "low".

We remark that [12] proves that every truthful-in-expectation mechanism can be replaced by an (approximately) maximal-in-distributional-range mechanism. However, this reduction involves a non-uniform step that causes problems in the setting of computational complexity, because [12] does not give any bound on the size of the non-uniform advice that would be needed. Therefore, we deal with truthful-in-expectation mechanisms directly here, without appealing to this reduction.

*Proof.* Assume that $\mathcal{M}$ is a truthful-in-expectation mechanism for the problem $\max\{f(S) : |S| = k\}$, where $f$ is a coverage function. Let $m$ denote the size of the ground set. First consider what happens when we run the mechanism on the input $f(S) = |S \cap A|$ where $A$ is some set of size $k$ ($f$ can be easily represented as a coverage function). $\mathcal{M}$ returns a distribution $D_A$ over sets of size $k$, and a price $p_A \geq 0$. Due to the approximation guarantee,
$$\mathbf{E}_{R \sim D_A}[|R \cap A|] \geq ck.$$
We also note that the price $p_A$ returned by the mechanism should have bit representation polynomial in $m$, since the polynomial is polynomial-time. We denote by
$$p_m = \frac{1}{\binom{m}{k}} \sum_{|A|=k} p_A$$
the average price over all sets $A$ of size $k$. This number still has a bit representation polynomial in $n$.

Now we describe how to solve the Max $k$-cover decision problem for $\epsilon = \Omega(c^2)$, using the mechanism $\mathcal{M}$ (and polynomial advice, namely the number $p_n$). Consider an instance of the Max $k$-cover problem on universe $U$, with the sets indexed by $M$ and denoted by $S_e, e \in M$. By examining Feige's reduction [14], we can see that in the YES case, there are $k$ sets covering the entire universe. Moreover, each element of the universe is covered by the same number of sets in the instance; let's call it $d$. The total number of sets is $dk$.

Let $\pi : M \to M$ be a random permutation of the set $M$. We let $f(T) = p_m |\bigcup_{e \in \pi(T)} S_e|$ be the input valuation to the mechanism. We claim the following: If we start from a YES instance, then the mechanism returns a distribution $D$ such that $\mathbf{E}_{R \sim D}[f(R)] \geq (1 - 1/e + 2\epsilon) p_m |U|$. If we start from a NO instance, then the mechanism returns a distribution $D$ such that $\mathbf{E}_{R \sim D}[f(R)] \leq (1 - 1/e + \epsilon) p_m |U|$. We can distinguish these two cases probabilistically by running the mechanism repeatedly, with the advice $p_m$. This would imply that $NP \subseteq BPP/poly = P/poly$.

In the case of NO instance, the proof is simple: There is no collection of $k$ sets covering more than a $(1 - 1/e + \epsilon)$-fraction of the universe. This means that $f(T) \leq (1 - 1/e + \epsilon) p_m |U|$ for every $|T| = k$, in particular for any solution returned by the mechanism.

Now suppose the Max $k$-cover instance is a YES instance. In this case there is a choice of $k$ sets that cover the universe, i.e. a set $A \subset M$ such that $f(\pi^{-1}(A)) = p_m |U|$. We feed $f$ as input into the mechanism. Recall that we shuffle the elements by a random permutation $\pi$, so the actual location of the optimal set is random from the point of view of the mechanism. However, for any particular set $A', |A'| = k$, there is a distribution $D_{A'}$ possibly returned by the mechanism that takes at least a $c$-fraction of $A'$ on average. We can condition on $\pi(A) = A'$, and the elements outside of $A'$ are still randomly shuffled. WLOG we can actually assume $A' = A$. Let us compute what expected value this distribution gives in terms of the objective function $f$:

$$\mathbf{E}_{R \sim D(A)}[f(R)] = p_m \mathbf{E}_{R \sim D(A)}[\bigcup_{e \in \pi(R)} S_e].$$

Note that $R$ takes at least $ck$ elements from $A$, and the remaining elements are taken randomly from the complement of $A$. In terms of the Max $k$-cover instance, this means that we take $c'k$ sets from the optimal solution ($\mathbf{E}[c'] \geq c$), and $(1-c')k$ sets from outside of $A$. Condition also on the number $c'$. Due to the random shuffle, every choice of $(1-c')k$ sets from outside of $A$ has the same probability. Since the total number of sets is $dk$, this means that each set outside of $A$ is taken with probability $(1-c)/d$. The negative correlations in the appearance of different sets can only help in terms of coverage probability; hence we can assume that the sets appear independently with probability $(1-c')/d$. Since each element not covered by $\bigcup_{e \in A} S_e$ appears in $d-1$ sets outside of $S$, this means that $R$ covers it with probability at least $1 - (1-(1-c')/d)^{d-1} \simeq 1 - e^{c'-1}$ (since $d$ will be a very large constant in the reduction). Therefore, we obtain

$$\mathbf{E}_{R \sim D(A)}[f(R)] \geq p_m(c' + (1-c')(1-e^{c'-1}))|U|.$$

This is a convex function of $c'$ which is equal to $(1-1/e)p_m|U|$ at $c' = 0$ and equal to $p_m|U|$ at $c' = 1$. For small $c' > 0$, it behaves like $(1 - 1/e + \Theta(c'^2))p_m|U|$. Since $\mathbf{E}[c'] \geq c$, due to convexity the worst case occurs when $c'$ is actually deterministic and equal to $c$. Therefore, for $\epsilon = \Theta(c^2)$ with a suitable constant, we will get $\mathbf{E}[f(R)] \geq (1 - 1/e + 3\epsilon)p_m|U|$.

By averaging over the random permutations $\pi$, we still get that the mechanism returns a solution of expected value at least $(1 - 1/e + 3\epsilon)p_m|U|$, and the average price is $p_m$. Therefore, the average utility (= expected value minus price) of this solution is at least $(1 - 1/e + 3\epsilon)p_m|U| - p_m \geq (1 - 1/e + 2\epsilon)p_m|U|$ (since $|U| \to \infty$).

The final argument is as follows. If there is a distribution of expected utility at least $(1 - 1/e + 2\epsilon)p_m|U|$ that the mechanism could possibly return on a certain input, by truthfulness in expectation this means that on the true input the mechanism must return a solution of expected utility at least $(1 - 1/e + 2\epsilon)p_m|U|$. The price cannot be negative due to individual rationality, hence the expected value of the returned solution must be at least $(1 - 1/e + 2\epsilon)p_m|U|$. This concludes the proof. □

## 5 Mechanisms for Multi-Unit Auctions

In a multi-unit auctions we have $m$ identical items and $n$ bidders. Each bidder $i$ has a valuation function $v_i : [m] \to \mathbb{R}$. We use the standard assumptions that the valuations are monotone and that $v_i(0) = 0$. The goal is to approximate the social welfare in time that is polynomial in $n$ and $\log m$. In [5] a truthful-in-expectation FPTAS for this problem was given, by constructing a maximal in distributional range algorithm. Furthermore, they show that if we there are two bidders and we require that the mechanism either always allocate all items or not allocate any item at all, then no universally truthful mechanism can provide an approximation ratio better than 2, but a simple adaption of the MIDR FPTAS is still a truthful-in-expectation FPTAS even for this restricted setting. The following theorem shows that the option to sometimes not allocate any item at all is crucial for obtaining good approximations using MIDR algorithms.

**Theorem 5.1.** *Let A be an MIDR algorithm for multi-unit auctions with two bidders that always allocates all items and runs in time $\mathrm{poly}(\log m)$. Then the approximation ratio of A is no better than $\frac{1}{2} + \epsilon$, unless $RP = NP$.*

*Proof.* We consider the following class of valuations that contains two types of valuations. The first type is single minded valuations: a valuation $v$ is called single minded when there exists some number $r$ such that for every $r' \geq r$ we have that $v(r') = 1$, and for every $r' < r$ we have that $v(r') = 0$. The second type contains a valuation $v_\phi$ for every SAT formula $\phi$ with $\log m$ variables: $v_f(s) = 2 \cdot s + \phi(s)$. That is, the value for the bundle of $s$ items is $2 \cdot s$, and a "bonus" of 1 if $s$ (viewed as a $\{0,1\}^{\log m}$ string and interpreted as an assignment to the variables) satisfies $\phi$. Notice that both types of valuations can be easily described in $O(\log m)$ space in the natural way.

**Lemma 5.2.** *Let $(x, m-x)$ be an allocation. Then, there is a distribution $D$ in the range such that $D$ outputs $(x, m-x)$ with probability at least $2\epsilon$.*

*Proof.* Consider the instance where both players are single minded as follows: Bidder 1 has a value of 1 for $x$ items or more and 0 for less than $x$ items, and Bidder 2 has a value of 1 for $m-x$ items or more otherwise his value is zero. Observe that all allocations except $(x, m-x)$ has a welfare of 1, and that $(x, m-x)$ has a welfare of 2. So to obtain an approximation ratio of at least $\frac{1}{2} + \epsilon$ (that is, to output a distribution with an expected welfare of $1 + 2\epsilon$), the distribution that $A$ outputs must have $(x, m-x)$ in its support with probability at least $2\epsilon$. □

Consider the instance where one bidder has a valuation $v_\phi$, for some $\phi$, and the other one has a valuation $u_{\phi'}$ where $\phi'$ is not satisfied by any assignment, i.e., $u(s) = 2 \cdot s$. Suppose that $\phi$ has a satisfying assignment $x$. Notice that every allocation that corresponds to a satisfying assumption has a welfare of $2m + 1$ and all other allocations have welfare of $2m$ (this is where the irremovable assumption that all items are allocated). From that and from the lemma, we have that there exists a distribution in the range that has an expected value of at least $2m + 2\epsilon$, and therefore the MIDR mechanism must return a distribution with at least that value. This implies that in the distribution that the algorithm outputs an allocation that corresponds to a satisfying assignment is returned with probability at least $2\epsilon$. Now we can run $A$ several times until it outputs an allocation with value $2m + 1$ and find a satisfying assignment to the SAT formula $\phi$. □

# A  Appendix for Section 3

## A.1  The Basic Instance

Here we describe the basic instance which is the starting point of our proof. This is identical to the basic instance in [13] and hence we only summarize the definition and state the lemma that we need without proof. Note that the parameters are somewhat different from [13], because of considerations arising in the next section.

**Basic instance**  We construct instances with $n$ players and $m$ items, where $n = 2^\ell$, $m = m_0 2^\ell$ ($m_0 \to \infty$ and $\ell$ is constant). In the *basic random instance*, player $i$ has a polar additive valuation $v_i(S) = v_{A_i^{(0)},\omega}(S) = |S \cap A_i^{(0)}| + \omega |S \setminus A_i^{(0)}|$, where $A_i^{(0)}$ is a uniformly random set of $m_0 = m/n$ items, chosen independently for each player. The following lemma is proved in [13].

**Lemma A.1.** *For any c-approximation mechanism applied to the basic random instance, there is a player $i$ and sets $A_j^{(0)}, j \neq i$, such that conditioned on the desired sets for players $j \neq i$ being $A_j^{(0)}$, player $i$ gets allocated a random set $R_i^{(0)}$ such that*

$$\mathbf{E}[|R_i^{(0)} \cap A_i^{(0)}|] > (c/4 - \omega)\mathbf{E}[|R_i^{(0)} \cup A_i^{(0)}|].$$

## A.2 Setup for Higher-Level Valuations

In the following, the valuations of all players except $i$ are fixed to be $v_j = v_{A_j^{(0)}, \omega}$ for the choice of sets $A_j^{(0)}, j \neq i$ given by Lemma A.1, and a parameter $\omega > 0$ to be fixed later. We will vary only the valuation of player $i$ who will be referred to as the *special player*. Since we work only with player $i$, we drop the index $i$ in the following. We consider the following random sequence of sets of items.

**Definition A.2.** *A random bisection sequence of a random sequence of pairs of sets $(A^{(0)}, B^{(0)})$, $(A^{(1)}, B^{(1)})$, ..., $(A^{(\ell)}, B^{(\ell)})$, generated as follows. We define $A^{(\ell)} = B^{(\ell)} = M$, the set of all items. Given $A^{(j)}$ for $0 < j \leq \ell$, we generate $(A^{(j-1)}, B^{(j-1)})$ uniformly at random among all partitions of $A^{(j)}$ into two parts of equal size.*

Observe that for each $j$, $(A^{(j)}, B^{(j)})$ is just a pair of random disjoint sets of size $2^{j-\ell} m$. However, the correlation between sets at different levels will be also important. We refer to $A^{(j)} = A^{(j-1)} \cup B^{(j-1)}$ as the $j$-th level of the bisection sequence.

For each pair of sets $(A^{(j-1)}, B^{(j-1)})$ and parameters $\alpha, \beta > 0$, we consider the valuation function $f_{A^{(j-1)}, B^{(j-1)}, \alpha, \beta}$, encoded on the input as $v_{A^{(j)}, \phi, \alpha, \beta}$ for some particular ordering of $A^{(j)}$ (see Section 3.1). Recall that this is the double-peak submodular function where $A^{(j-1)}$ and $B^{(j-1)}$ are the two desired sets of items.

**Density menu** Recall that according to the taxation principle, if the special player $i$ reports valuation $v_i$, the mechanism must allocate to him a distribution over sets and prices of optimal expected utility out of all the distributions potentially allocated under some reported valuation. The menu is the collection of all such possible distributions (given fixed valuations for all other players). Similarly to [13], we replace the menu here by an object that instead of the full distributions contains only the information relevant to us, namely the "density" of a random set from the distribution, restricted to the set of relevant items $A^{(j)}$. We call this object the "density menu".

**Definition A.3.** *Given a mechanism, a special player, and fixed valuations for the other players, the "density menu of level $j$", $\mathcal{M}_j$, is a collection of probability distributions over $\mathbb{R}^2$ (pairs of real numbers), defined as follows: The distribution of a pair of random variables $(X_j, P_j)$ is in $\mathcal{M}_j$, if $(X_j, P_j)$ can be generated as*

$$X_j = \frac{|A^{(j)} \cap R(A^{(j)})|}{|A^{(j)}|},$$

$$P_j = P(A^{(j)}),$$

*where $A^{(j)}$ is a uniformly random set of $2^{j-\ell} m$ items, and $(R(A), P(A))$ is a random (set, price) pair allocated by the mechanism to the special player for some reported valuation $v_A$ depending on $A$.*

In other words, $X_j$ is the (random) fraction of the desired items that the special player can potentially receive at level $j$, and $P_j$ is the respective price. Note that there are two sources of randomness in $(X_j, P_j)$: the random choice of $A^{(j)}$, and the randomness in the distribution returned by the mechanism for a fixed valuation $v_A$. We illustrate this by an example:

- Consider $m = 4$ and $\ell = 1$; $A^{(0)}$ is a uniformly random set of 2 elements out of 4.
- Suppose that for each pair $A$, there is a valuation $v_A'$ such that the special player reporting $v_A'$ receives a (deterministic) set $R(A)$ such that $|A \cap R(A)| = 1$, at a price of 500. This means that $\mathcal{M}_0$ contains a distribution such that $(X_0, P_0) = (1/2, 500)$ with probability 1.
- Suppose that for another choice of $v_A''$ for each pair $A$, the special player receives a random set $R(A)$ such that $|A \cap R(A)| = 0$ and the price is 0, or $|A \cap R(A)| = 2$ and the price is 1000, each with prob. $1/2$. Then $\mathcal{M}_0$ contains a distribution such that $(X_0, P_0) = (0, 0)$ or $(X_0, P_0) = (1, 1000)$, each with prob. $1/2$.

- Finally, consider the choice of $v'_A$ for half of the choices of $A$, and $v''_A$ for another half of the choices of $A$. Whenever reporting $v'_A$, the player receives 1 element of $A$ at a price of 500. When reporting $v''_A$, he receives 0 or 2 with probability $1/2$, at a price of 0 or 1000. Overall, he receives 0 elements from $A$ with probability $1/4$, 1 element with probability $1/2$, and 2 elements with probability $1/4$. Hence, $\mathcal{M}_0$ also contains the distribution $(X_0, P_0) = (0,0)$ with probability $1/4$, $(1, 500)$ with probability $1/2$ and $(2, 1000)$ with probability $1/4$.

**Closure of a density menu** Furthermore, it will be convenient to make the menu closed and convex as follows.

**Definition A.4.** *We define $\overline{\mathcal{M}_j}$, the closure of the density menu at level $j$, to be the topological closure of the set of all convex combinations of distributions from the menu $\mathcal{M}_j$.*

To avoid confusion, we emphasize that the convex combinations occur in the space of distributions, i.e. they are obtained by averaging *probabilities* of each possible outcome, and not the *values* of $(X_j, P_j)$. A distribution of $(X_j, P_j)$ is in $\overline{\mathcal{M}_j}$, if its distribution can be approximated arbitrarily closely by some convex combination of distributions in $\mathcal{M}_j$.

## A.3 The Main Reduction

The oracle hardness proof in [13] relies on the fact that a mechanism bounded by a polynomial number of value queries to a valuation $f_{A,B,\alpha,\beta}$ cannot identify the two sets $A, B$ and with high probability returns a distribution that is balanced with respect to these two sets. Then, this argument is harnessed to show that at each successive level, the mechanism must contain distributions in the menu that are in some sense more valuable than at the previous level. We call this the *density-boosting property*. Eventually, the density-boosting property leads to a contradiction with truthfulness at the last level.

Here, we use this argument differently: We argue that unless a contradiction with truthfulness occurs at the last level, there must be a pair of successive levels where the density-boosting property is violated. This allows us to find parameters for embedding the Unique-SAT problem in our instance and solving it using the presumed mechanism. The fact that the density-boosting property is violated at some level follows directly from [13]; but we need to prove a little bit more, namely that the parameters describing the violation can be encoded succinctly by polynomially many bits. The precise statement that we want to prove is the following.

**Definition A.5.** *For a parameter $\alpha > 0$, define $\psi_\alpha(x) = \min\{x/\alpha, 1\}$.*

**Lemma A.6.** *There are absolute constants $\gamma, \epsilon > 0$ such that the following holds. Consider a $(1-\epsilon)$-approximately truthful-in-expectation mechanism for combinatorial auctions with submodular valuations, $n = 2^\ell$ players and $m = m_0 2^\ell$ items (for some constant $\ell$ and arbitrary $m_0$), providing a $c$-approximation in social welfare for $c = 2^{-\gamma \ell}$. Then there is $j \in \{0, \ldots, \ell - 1\}$ and parameters $\alpha > 0$, $\lambda \in [\frac{1}{2\ell}, 10^\ell(m+2)]$, described by polynomially many bits such that there is a pair of random variables $(X_j, P_j)$ with a distribution in the menu closure $\overline{\mathcal{M}_j}$, and for every $(X_{j+1}, P_{j+1})$ with distribution in $\overline{\mathcal{M}_{j+1}}$, we have*

$$\mathbf{E}[\lambda(1 - (1 - \psi_\alpha(X_{j+1}))^2) - P_{j+1}] + \frac{1}{4\ell} < (1-\epsilon)\mathbf{E}[\lambda \psi_\alpha(X_j - 10^{-\ell})] - \mathbf{E}[P_j].$$

Assuming this lemma, we finish the proof of Theorem 3.2 as follows.

*of Theorem 3.2.* Assume that there is a $(1 - \epsilon)$-approximately truthful-in-expectation $n^{-\gamma}$-approximation for combinatorial auctions with $n$ bidders and valuations of the type described above (polar and double-peak valuations). We can assume w.l.o.g. that $n = 2^\ell$ (otherwise let us prove the theorem for the nearest power of 2, and adjust $\gamma$ accordingly). We are given a Unique-SAT formula $\phi$ on $m'$ variables and we generate an instance of combinatorial auctions that will enable us to solve the Unique-SAT instance. We also describe an advice string depending only on the size of the formula, which will be needed to complete the reduction.

We set $\beta = 10^{-\ell}$ and $m = nm'' = 2^\ell m''$, where $m'' = poly(m')$ are the parameters of an $(m'', m', \frac{1}{2}(1-\beta)m'')$-list-decodable code. Let $j$ be the level provided by Lemma A.6 for $n$ players and $m$ items. If $j > 0$, we adjust the code by duplicating each bit of a codeword $2^j$ times, so that codewords have length $2^{j-\ell}m$. The double-peak valuations using this code are then defined on a support of size $2^{j+1-\ell}m$, which is what we need at level $j$.

For players $i' \neq i$, let their true valuations be the polar valuations $v_{A_{i'},\omega}, i' \neq i$ guaranteed by Lemma A.1. (These sets depend only on the size of the instance and hence can be given as part of the advice string.) For player $i$, let the true valuation be $\lambda v_{C,\phi,\alpha,\beta}$ where $C$ is a randomly ordered random set of $2m'' = 2^{j-\ell}m$ items, and $j, \alpha, \lambda$ are the parameters given by Lemma A.6. Again, these parameters as well as $\beta$ are part of the advice string. ($\lambda v$ denotes multiplication of a valuation function pointwise by a scalar $\lambda$.) We run the mechanism with $C$ and its ordering randomly chosen, and all calculations will be in expectation over this random choice.

Now, if $\phi$ is a satisfiable formula, we denote by $(A, B)$ the partition of $C$ corresponding to the (unique) satisfiable assignment. Then the valuation is $\lambda v_{C,\phi,\alpha,\beta} = \lambda f_{A,B,\alpha,\beta}$. Recall the definition of $f_{A,B,\alpha,\beta}(S) = \tilde{\psi}(x, y)$ from Section 3.1, using the variables $x = \frac{|S \cap A|}{|A|}$ and $y = \frac{|S \cap B|}{|B|}$, and observe that in all cases,

$$\lambda v_{C,\phi,\alpha,\beta}(S) = \lambda \tilde{\psi}(x, y) \geq \lambda \tilde{\psi}(x, 0) \geq \lambda \min\left\{\frac{x - \beta}{\alpha}, 1\right\} = \lambda \psi_\alpha(x - \beta).$$

Due to $C$ and its ordering being chosen randomly, $(A, B)$ is distributed uniformly over all pairs of disjoint sets of size $2^{j-\ell}m$. In particular, the set $A$ is a uniformly random set of size $2^{j-\ell}m$. By the definition of a density menu $\mathcal{M}_j$ (Definition A.3), if a pair of random variables $(X_j, P_j)$ has a distribution in $\overline{\mathcal{M}_j}$, it means that there are valuations $v_A$ that player $i$ could declare, depending on $A$, such that the fraction of $A$ that he receives is given by the random variable $X_j$, and the price he is charged is $P_j$. Considering the above description of player $i$'s true valuation, where the fraction of $A$ received is denoted by $x = \frac{|S \cap A|}{|A|}$, the respective utility that the special player obtains from such a distribution is at least

$$\mathcal{U} = \mathbf{E}[\lambda \psi_\alpha(X_j - \beta) - P_j] = \mathbf{E}[\lambda \psi_\alpha(X_j - 10^{-\ell}) - P_j].$$

We remark that since this expected utility is achieved by a distribution in the menu closure $\overline{\mathcal{M}_j}$, at least the same expected utility (or arbitrarily close) can be achieved also by a distribution on the menu $\mathcal{M}_j$ itself (using the fact that taking convex combinations of distributions of $X_j$ results in convex combinations of expected values, of any function of $X_j$). By the property of approximate truthfulness in expectation, if utility at least $\mathcal{U}$ can be achieved for some valuation possibly declared by the player, then utility at least $(1 - \epsilon)\mathcal{U} \geq (1 - \epsilon)\mathbf{E}[\lambda \psi_\alpha(X_j - 10^{-\ell})] - \mathbf{E}[P_j]$ must be achieved when declaring the true valuation $\lambda v_{C,\phi,\alpha,\beta}$. Observe that this is the right-hand side of the inequality in Lemma A.6.

Now consider the random set $R$ that the player $i$ actually receives when declaring the true valuation $\lambda v_{C,\phi,\alpha,\beta}$. We claim that this set must be at least sometimes significantly unbalanced with respect to the partition $(A, B)$. Namely, we claim

$$\Pr\left[\frac{|R \cap A|}{|A|} - \frac{|R \cap B|}{|B|} \in [-\beta, +\beta]\right] < 1 - \frac{1}{4\ell\lambda}. \tag{1}$$

Assume for a contradiction that $\frac{|R \cap A|}{|A|} - \frac{|R \cap B|}{|B|} \in [-\beta, +\beta]$ with probability at least $1 - \frac{1}{4\ell\lambda}$. As before, we denote $x = \frac{|R \cap A|}{|A|}$ and $y = \frac{|R \cap B|}{|B|}$. When $x - y \in [-\beta, +\beta]$, we have by the construction of the valuation $v_{C,\phi,\alpha,\beta}$,

$$v_{C,\phi,\alpha,\beta}(R) = 1 - \left(1 - \frac{x+y}{2\alpha}\right)_+^2 = 1 - \left(1 - \frac{|R \cap C|}{\alpha|C|}\right)_+^2 = 1 - (1 - \psi_\alpha(X_{j+1}))^2$$

where $X_{j+1} = \frac{|R \cap C|}{|C|}$. When $|x - y| > \beta$, the value $v_{C,\phi,\alpha,\beta}(R)$ can increase, compared to this formula, by no more than 1. Since this happens with probability at most $\frac{1}{4\ell\lambda}$, we get

$$\mathbf{E}[v_{C,\phi,\alpha,\beta}(R)] \leq \mathbf{E}[1 - (1 - \psi_\alpha(X_{j+1}))^2] + \frac{1}{4\ell\lambda}.$$

The expectation here is over $C$, its ordering and the randomness in $R$. By the definition of level-$(j+1)$ menu (Definition A.3), the menu $\mathcal{M}_{j+1}$ contains the distribution of the pair $(X_{j+1}, P_{j+1})$ where $X_{j+1} = \frac{|R \cap C|}{|C|}$ and $P_{j+1}$ is the respective random price that player $i$ is charged. By the above inequality, the utility of player $i$ when declaring his true valuation is

$$\mathbf{E}[\lambda v_{C,\phi,\alpha,\beta}(R) - P_{j+1}] \leq \lambda \mathbf{E}[\lambda(1 - (1 - \psi_\alpha(X_{j+1}))^2) - P_{j+1}] + \frac{1}{4\ell}.$$

However, Lemma A.6 states that for any distribution in $\mathcal{M}_{j+1}$ (even in $\overline{\mathcal{M}_{j+1}}$),

$$\mathbf{E}[\lambda(1 - (1 - \psi_\alpha(X_{j+1}))^2) - P_{j+1}] + \frac{1}{4\ell} < (1-\epsilon)\mathbf{E}[\lambda \psi_\alpha(X_j - 10^{-\ell})] - \mathbf{E}[P_j] \leq (1-\epsilon)\mathcal{U}.$$

Recall we proved that the right-hand side is a lower bound on the utility that player $i$ receives when declaring his true valuation. This is a contradiction which proves (1).

Therefore, with probability at least $\frac{1}{4\ell\lambda}$, the returned set $R$ satisfies $\frac{|R \cap A|}{|A|} - \frac{|R \cap B|}{|B|} \notin [-\beta, +\beta]$. But once we find such a set (by running the mechanism polynomially many times, this will happen with high probability), we can determine the partition of $C$ into $(A, B)$ by decoding the set $R \cap C$ using the list-decodable code, and checking whether any of the decoded strings $x$ is a satisfying assignment to the formula $\phi$. In conclusion, we solve the Unique-SAT problem and hence also the SAT problem (by a randomized reduction) in $RP/poly = P/poly$. □

## A.4 Proof of Lemma A.6

In this section, we prove Lemma A.6, which completes the proof. This lemma relies on the density-boosting technique of [13]. In particular, we appeal to the following technical lemma proved in [13]. The content of this lemma is a boosting argument about distributions at different levels that leads to an exponential blow-up in terms of density.

**Lemma A.7.** *There are absolute constants $\epsilon, \delta > 0$ such that the following holds for any sufficiently large $\ell \in \mathbb{N}$. If $\mathcal{X}_0, \ldots, \mathcal{X}_\ell$ are collections of random variables in $[0,1]$ such that*

- *there is $X_0$ in $\mathcal{X}_0$ such that $\mathbf{E}[X_0] \geq c$ for some $c \geq 2^{-\ell}$, and*
- *for every $X_j$ in $\mathcal{X}_j$ and for every function of the form $\psi_\alpha(t) = \min\left\{\frac{t}{\alpha}, 1\right\}$, $\alpha \in [\delta^{\ell/2}, 1]$, there is $X_{j+1}$ in $\mathcal{X}_{j+1}$ such that*

$$\mathbf{E}[1 - (1 - \psi_\alpha(X_{j+1}))^2] \geq (1-\epsilon)\mathbf{E}[\psi_\alpha(X_j - 10^{-\ell})] - 10^{-\ell}$$

*then there is a sequence of variables $X_j$ in $\mathcal{X}_j$ and parameters $1 = \alpha_0 \geq \alpha_1 \geq \ldots \alpha_\ell \geq \delta^{\ell/2}$ such that for each $j = 1, 2, \ldots, \ell$,*

$$\alpha_j (\mathbf{E}[\psi_{\alpha_j}(X_j)])^{1+\delta} \geq \left(\frac{1+\delta^2}{2}\right)^j c^{1+\delta}.$$

We remark that instead of "for every function of the form $\psi_\alpha(t) = \min\left\{\frac{t}{\alpha}, 1\right\}$", the assumption in [13] is formulated as "for every non-decreasing concave function". However, the only non-decreasing concave functions that are used in the proof are of the form above. Also, the possible parameters $\alpha$ arising in the proof are in the interval $[\delta^{\ell/2}, 1]$, because $\alpha_0 = 1$ and in each step $\alpha$ can decrease by at most a factor of $\sqrt{\delta}$. Therefore Lemma A.7 follows from the proof in [13].

Using this lemma, we first prove the following as a stepping stone towards Lemma A.6.

**Lemma A.8.** *There are absolute constants $\gamma, \delta, \epsilon > 0$ such that the following holds. Consider a $(1-\epsilon)$-approximately truthful-in-expectation mechanism for combinatorial auctions with submodular valuations, $n = 2^\ell$ players and $m = m_0 2^\ell$ items (for some constant $\ell$ and arbitrary $m_0$), providing a $c$-approximation in social* welfare for $c = 2^{-\gamma\ell}$. Then there is $j \in \{0, \ldots, \ell-1\}$ and a parameter $\alpha \in [\delta^{\ell/2}, 1]$ described by polynomially many bits such that the menu closure $\overline{\mathcal{M}_j}$ contains a distribution of a pair $(X_j, P_j)$ such that $\mathbf{E}[P_j] \leq m+1$, and for every $(X_{j+1}, P_{j+1})$ with a distribution in $\overline{\mathcal{M}_{j+1}}$, either

$$\mathbf{E}[1 - (1 - \psi_\alpha(X_{j+1}))^2] < (1-\epsilon)\mathbf{E}[\psi_\alpha(X_j - 10^{-\ell})] - 10^{-\ell}$$

or

$$\mathbf{E}[P_{j+1}] > \mathbf{E}[P_j] + \frac{1}{\ell}.$$

*Proof.* Let $\epsilon, \delta > 0$ be the constants provided by Lemma A.7. We define $\gamma = \frac{\delta^2}{2(1+\delta)}$. Assume to the contrary that we have a $(1-\epsilon)$-approximately truthful-in-expectation mechanism for $n = 2^\ell$ players, $m = m_0 2^\ell$ items, providing a $c = 2^{-\gamma\ell}$-approximation in social welfare. In addition, for every $j \in \{0, \ldots, \ell-1\}$, $\alpha \in [\delta^{\ell/2}, 1]$ and every distribution $(X_j, P_j)$ in $\overline{\mathcal{M}_j}$, with $\mathbf{E}[P_j] \leq m+1$, there is a distribution $(X_{j+1}, P_{j+1})$ in $\overline{\mathcal{M}_{j+1}}$ such that

$$\mathbf{E}[1 - (1 - \psi_\alpha(X_{j+1}))^2] \geq (1-\epsilon)\mathbf{E}[\psi_\alpha(X_j - 10^{-\ell})] - 10^{-\ell}$$

and

$$\mathbf{E}[P_{j+1}] \leq \mathbf{E}[P_j] + \frac{1}{\ell}.$$

We start from the basic instance (Section A.1), where we set $\omega = c/8$. According to Lemma A.1, we choose a special player and fix the valuations of the other players. Let $R^{(0)}$ be the random set allocated to the special player, $P_0$ the respective price, $A^{(0)}$ his desired set, $X_0 = \frac{|R^{(0)} \cap A^{(0)}|}{|A^{(0)}|}$, $c_0 = \mathbf{E}[X_0]$ and $p_0 = \mathbf{E}[P_0]$. Lemma A.1 implies $c_0 \geq c/4 - \omega = c/8 = 2^{-\gamma\ell-3}$. Observe also that $p_0$ cannot be very large due to individual rationality: valuations in the basic instance are bounded by $m$, so we have $p_0 \leq m$.

Now let us consider the density menu at different levels and their closures $\overline{\mathcal{M}_j}$ (Section A.2). We define $\mathcal{X}_j$ to be the collection of random variables $X_j$ such that $(X_j, P_j)$ is in $\overline{\mathcal{M}_j}$ for some price $P_j$ such that $\mathbf{E}[P_j] \leq p_0 + \frac{j}{\ell}$. As discussed above, we have $X_0$ in $\mathcal{X}_0$ such that $\mathbf{E}[X_0] = c_0 \geq 2^{-\gamma\ell-3} \geq 2^{-\ell}$ for $\gamma\ell$ sufficiently large (which can be assumed, because this corresponds to the approximation factor $c = 2^{-\gamma\ell}$ being sufficiently small).

By our assumption by contradiction (beginning of proof), the collections $\mathcal{X}_0, \mathcal{X}_1, \ldots, \mathcal{X}_\ell$ satisfy the assumptions of the technical Lemma A.7. (Note that prices can increase by $1/\ell$ in each step, which is consistent with our definition of $\mathcal{X}_j$.) Consequently, the lemma implies that there is a random variable $X_\ell$ in the collection $\mathcal{X}_\ell$ and $\alpha_\ell \in [\delta^{\ell/2}, 1]$ such that

$$\alpha_\ell(\mathbf{E}[\psi_{\alpha_\ell}(X_\ell)])^{1+\delta} \geq \left(\frac{1+\delta^2}{2}\right)^\ell c_0^{1+\delta}.$$

Recall that $\psi_{\alpha_\ell}(t) = \min\{\frac{t}{\alpha_\ell}, 1\}$. Therefore,

$$\mathbf{E}[X_\ell] \geq \alpha_\ell \mathbf{E}[\psi_{\alpha_\ell}(X_\ell)] \geq \alpha(\mathbf{E}[\psi_{\alpha_\ell}(X_\ell)])^{1+\delta} \geq \left(\frac{1+\delta^2}{2}\right)^\ell c_0^{1+\delta} \geq 2^{\delta^2\ell - \ell} c_0^{1+\delta}.$$

Since $X_\ell$ is in $\mathcal{X}_\ell$, the respective price is bounded by $\mathbf{E}[P_\ell] \leq p_0 + 1$. Now consider the following expression:

$$\mathbf{E}[\omega m X_\ell - P_\ell] \geq \omega m 2^{\delta^2\ell - \ell} c_0^{1+\delta} - p_0 - 1.$$

The distribution of $(X_\ell, P_\ell)$ is in the menu closure $\overline{\mathcal{M}_\ell}$; recall that this involves taking convex combinations and limits of distributions on the actual menu $\mathcal{M}_\ell$. Therefore, a slightly weaker linear equality must be satisfied by some pair $(\tilde{X}_\ell, \tilde{P}_\ell)$ with a distribution on the actual menu $\mathcal{M}_\ell$:

$$\mathbf{E}[\omega m \tilde{X}_\ell - \tilde{P}_\ell] \geq \omega m 2^{\delta^2\ell - \ell} c_0^{1+\delta} - p_0 - 2.$$

Recall that being on the menu $\mathcal{M}_\ell$ means that $\tilde{X}_\ell = \frac{|\tilde{R}^{(\ell)}|}{|A^{(\ell)}|} = \frac{1}{m}|\tilde{R}^{(\ell)}|$, where $\tilde{R}^{(\ell)}$ is a random set allocated to the special player for a certain valuation at level $\ell$. Consider now what would happen if the special player reports this level-$\ell$ valuation in the basic instance (where his true valuation is a polar one, $v_{A^{(0)},\omega}(S) = |S \cap A_i| + \omega|S \setminus A_i|$). He would receive expected utility

$$\mathbf{E}[v_{A_i,\omega}(\tilde{R}^{(\ell)}) - \tilde{P}_\ell] \geq \mathbf{E}[\omega|\tilde{R}^{(\ell)}| - \tilde{P}_\ell] \geq \mathbf{E}[\omega m \tilde{X}_\ell - \tilde{P}_\ell] \geq \omega m 2^{\delta^2 \ell - \ell} c_0^{1+\delta} - p_0 - 2.$$

Using $c_0 \geq c/8 = 2^{-\gamma \ell - 3}$, we get

$$\mathbf{E}[v_{A^{(0)},\omega}(\tilde{R}^{(\ell)}) - \tilde{P}_\ell] \geq \omega m 2^{\delta^2 \ell - \ell} c_0^{1+\delta} - p_0 - 2 \geq 2^{\delta^2 \ell - (1+\delta)(\gamma \ell + 3)} m_0 c_0 - p_0 - 2.$$

Since we chose $\gamma = \frac{\delta^2}{2(1+\delta)}$, this means

$$\mathbf{E}[v_{A^{(0)},\omega}(\tilde{R}^{(\ell)}) - \tilde{P}_\ell] \geq 2^{\delta^2 \ell / 2 - 3(1+\delta)} m_0 c_0 - p_0 - 2. \tag{2}$$

On the other hand, the set $R^{(0)}$ actually allocated under declared valuation $v_{A^{(0)},\omega}$ gives

$$\mathbf{E}[v_{A^{(0)},\omega}(R^{(0)})] = \mathbf{E}[|R^{(0)} \cap A^{(0)}|] + \omega \mathbf{E}[|R^{(0)} \setminus A^{(0)}|] \leq \frac{m}{n}\mathbf{E}[X_0] + \omega \mathbf{E}[|R^{(0)}|] \leq \frac{2m}{n}\mathbf{E}[X_0]$$

using again Lemma A.1 to say that $\frac{m}{n}\mathbf{E}[X_0] = \mathbf{E}[|R^{(0)} \cap A^{(0)}|] \geq (c/4 - \omega)\mathbf{E}[|R^{(0)}|] = \omega \mathbf{E}[|R^{(0)}|]$. Therefore, since $\mathbf{E}[X_0] = c_0$ and $\mathbf{E}[P_0] = p_0$,

$$\mathbf{E}[v_{A^{(0)},\omega}(R^{(0)}) - P_0] \leq \frac{2m}{n}\mathbf{E}[X_0] - \mathbf{E}[P_0] = 2m_0 c_0 - p_0. \tag{3}$$

Comparing (2) and (3), we see that for sufficiently large $m_0$ and $\ell$, the special player is significantly better off reporting a level-$\ell$ valuation that gives him the random set $\tilde{R}^{(\ell)}$ at a price $\tilde{P}_\ell$, rather than reporting his true valuation $v_{A^{(0)},\omega}$. This would violate the assumption of $(1-\epsilon)$-approximate truthfulness in expectation. Therefore, we have proved the lemma by contradiction. $\square$

Now we are almost done - we just have to use Lemma A.8 in order to prove Lemma A.6. This is essentially a convex separation argument in the plane: we want to turn two separation inequalities for a convex set into one inequality. However, some technicalities arise because of the requirement that the separating inequality should be described by a polynomial number of bits.

*of Lemma A.6.* Assume that we have a mechanism as described in the lemma. These are the same conditions that we assume in Lemma A.8. Lemma A.8 implies that there is $j \in \{0, \ldots, \ell - 1\}$ and a parameter $\alpha \in [\delta^{\ell/2}, 1]$ described by polynomially many bits such that the menu closure $\overline{\mathcal{M}}_j$ contains a distribution $(X_j, P_j)$ such that for every distribution $(X_{j+1}, P_{j+1})$ in $\overline{\mathcal{M}}_{j+1}$ either

$$\mathbf{E}[1 - (1 - \psi_\alpha(X_{j+1}))^2] < (1-\epsilon)\mathbf{E}[\psi_\alpha(X_j - 10^{-\ell})] - 10^{-\ell}$$

or

$$\mathbf{E}[P_{j+1}] > \mathbf{E}[P_j] + \frac{1}{\ell}.$$

Define

$$\mathcal{Q}_{j+1} = \{(\mathbf{E}[1 - (1-\psi_\alpha(X_{j+1}))^2], \mathbf{E}[P_{j+1}]) : (X_{j+1}, P_{j+1}) \in \overline{\mathcal{M}}_j\}.$$

Then $\mathcal{Q}_{j+1}$ is a closed convex set in the plane, because taking convex combinations of distributions of $(X_{j+1}, P_{j+1})$ corresponds to taking convex combinations of the respective points in the plane. Moreover, by the above $\mathcal{Q}_{j+1}$ is disjoint from the closed convex set

$$\mathcal{R}_j = \{(q, p) : 1 \geq q \geq (1-\epsilon)\mathbf{E}[\psi_\alpha(X_j - 10^{-\ell})] - 10^{-\ell}, 0 \leq p \leq \mathbf{E}[P_j] + \frac{1}{\ell}\}.$$

Next, we replace $\mathcal{R}_j$ by a subset $\mathcal{R}'_j \subset \mathcal{R}_j$, which is defined as follows:

$$\mathcal{R}'_j = \{(q,p) \ : \ q \leq 1, p \geq 0,$$
$$\frac{1}{2\ell}(q - (1-\epsilon)\mathbf{E}[\psi_\alpha(X_j - 10^{-\ell})]) \geq p - \mathbf{E}[P_j] - \frac{1}{2\ell},$$
$$10^\ell(m+2)(q - (1-\epsilon)\mathbf{E}[\psi_\alpha(X_j - 10^{-\ell})]) \geq p - \mathbf{E}[P_j] - \frac{1}{2\ell}\}.$$

It can be verified that $\mathcal{R}'_j \subset \mathcal{R}_j$, using the bounds $q \leq 1$ and $\mathbf{E}[P_j] \leq m+1$: every point $(q,p)$ satisfying the constraints of $\mathcal{R}'_j$ also satisfies the constraints of $\mathcal{R}_j$.

Since $\mathcal{Q}_{j+1}$ is disjoint from $\mathcal{R}'_j$, they can be separated by a line through the point $((1-\epsilon)\mathbf{E}[\psi_\alpha(X_j - 10^{-\ell})], \mathbf{E}[P_j] + \frac{1}{2\ell})$ which is on the boundary of $\mathcal{R}'_j$: there is a coefficient $\lambda$ such that for each $(X_{j+1}, P_{j+1}) \in \overline{\mathcal{M}}_j$, we have

$$\lambda(\mathbf{E}[1 - (1-\psi_\alpha(X_{j+1}))^2] - (1-\epsilon)\mathbf{E}[\psi_\alpha(X_j - 10^{-\ell})]) < \mathbf{E}[P_{j+1}] - \mathbf{E}[P_j] - \frac{1}{2\ell}.$$

Moreover, the coefficient $\lambda$ must be in the interval $[\frac{1}{2\ell}, 10^\ell(m+2)]$, otherwise this constraint does not separate from $\mathcal{R}'_j$ (see the constraints defining $\mathcal{R}'_j$ above). Finally, we round $\lambda$ down to a multiple of $\frac{1}{4\ell}$, to obtain $\lambda' = \lfloor 4\ell\lambda \rfloor \frac{1}{4\ell}$. Since we have $|\lambda' - \lambda| \leq \frac{1}{4\ell}$, the following inequality is still valid:

$$\lambda'(\mathbf{E}[1 - (1-\psi_\alpha(X_{j+1}))^2] - (1-\epsilon)\mathbf{E}[\psi_\alpha(X_j - 10^{-\ell})]) < \mathbf{E}[P_{j+1}] - \mathbf{E}[P_j] - \frac{1}{4\ell}.$$

As $\lambda'$ is in the form $\lambda' = \frac{\ell'}{4\ell}$ where $\ell$ is a constant and $\ell'$ is an integer bounded by $4\ell \cdot 10^\ell(m+2)$, $\lambda'$ can be described by a polynomial number of bits. This is the inequality claimed by Lemma A.6. $\square$